\documentclass[english,preprint]{elsarticle}
\usepackage[T1]{fontenc}
\usepackage[latin9]{inputenc}
\usepackage{babel}
\usepackage{geometry}
\usepackage{array}
\usepackage{amssymb}
\usepackage{amsmath}
\usepackage{graphicx}
\usepackage[unicode=true,
 bookmarks=true,bookmarksnumbered=false,bookmarksopen=false,
 breaklinks=false,pdfborder={0 0 1},backref=false,colorlinks=false]
 {hyperref}
\usepackage{breakurl}

\makeatletter

\usepackage{sidecap} 

\usepackage{natbib}

\@ifundefined{showcaptionsetup}{}{%
 \PassOptionsToPackage{caption=false}{subfig}}
\usepackage{subfig}
\makeatother

\begin{document}

\title{An Ultra-Low Background PMT for Liquid Xenon Detectors}

\author[cwru]{D.S. Akerib}

\author[sdsmt]{X. Bai}

\author[yale]{E. Bernard}

\author[llnl]{A. Bernstein}

\author[cwru]{A. Bradley}

\author[usd]{D. Byram}

\author[yale]{S.B. Cahn}

\author[cwru]{M.C. Carmona-Benitez}

\author[llnl]{D. Carr}

\author[brown]{J.J. Chapman}

\author[cwru]{K. Clark}

\author[cwru]{T. Coffey}

\author[yale]{B. Edwards}

\author[coimbra]{L. de\,Viveiros}

\author[cwru]{M. Dragowsky}

\author[uor]{E. Druszkiewicz}

\author[brown]{C.H. Faham}

\author[brown]{S. Fiorucci}

\author[brown]{R.J. Gaitskell}

\author[cwru]{K.R. Gibson}

\author[umd]{C. Hall}

\author[sdsmt]{M. Hanhardt}

\author[davis]{B. Holbrook}

\author[ucb]{M. Ihm}

\author[ucb]{R.G. Jacobsen}

\author[yale]{L. Kastens}

\author[llnl]{K. Kazkaz}

\author[yale]{N. Larsen}

\author[cwru]{C. Lee}

\author[coimbra]{A. Lindote}

\author[coimbra]{M.I. Lopes}

\author[yale]{A. Lyashenko}

\author[brown]{D.C. Malling\corref{cor1}}
  
\cortext[cor1]{Corresponding Author. Tel.: +1 401 863 6222. E-mail address: David\_Malling@brown.edu}

\author[tamu]{R. Mannino}

\author[yale]{D.N. McKinsey}

\author[usd]{D.-M Mei}

\author[davis]{J. Mock}

\author[harvard]{M. Morii}

\author[ucsb]{H. Nelson}

\author[coimbra]{F. Neves}

\author[yale]{J.A. Nikkel}

\author[brown]{M. Pangilinan}

\author[cwru]{P. Phelps}

\author[cwru]{T. Shutt}

\author[coimbra]{C. Silva}

\author[uor]{W. Skulski}

\author[coimbra]{V.N. Solovov}

\author[llnl]{P. Sorensen}

\author[usd]{J. Spaans}

\author[tamu]{T. Stiegler}

\author[davis]{M. Sweany}

\author[davis]{M. Szydagis}

\author[ucb]{D. Taylor}

\author[davis]{J. Thomson}

\author[davis]{M. Tripathi}

\author[davis]{S. Uvarov}

\author[brown]{J.R. Verbus}

\author[davis]{N. Walsh}

\author[tamu]{R. Webb}

\author[tamu]{J.T. White}

\author[harvard]{M. Wlasenko}

\author[uor]{F.L.H. Wolfs}

\author[davis]{M. Woods}

\author[usd]{C. Zhang}

\address[brown]{Brown University, Dept. of Physics, 182 Hope St., Providence RI 02912, USA}

\address[cwru]{Case Western Reserve University, Dept. of Physics, 10900 Euclid Ave, Cleveland OH 44106, USA}

\address[harvard]{Harvard University, Dept. of Physics, 17 Oxford St., Cambridge MA 02138, USA}

\address[llnl]{Lawrence Livermore National Laboratory, 7000 East Ave., Livermore CA 94551, USA}

\address[coimbra]{LIP-Coimbra, Department of Physics, University of Coimbra, Rua Larga, 3004-516 Coimbra, Portugal}

\address[sdsmt]{South Dakota School of Mines and technology, 501 East St Joseph St., Rapid City SD 57701, USA}

\address[tamu]{Texas A \& M University, Dept. of Physics, College Station TX 77843, USA}

\address[ucb]{University of California Berkeley, Department of Physics, Berkeley, CA 94720-7300, USA}

\address[davis]{University of California Davis, Dept. of Physics, One Shields Ave., Davis CA 95616, USA}

\address[ucsb]{University of California Santa Barbara, Dept. of Physics, Santa Barbara CA 93106, USA}

\address[umd]{University of Maryland, Dept. of Physics, College Park MD 20742, USA}

\address[uor]{University of Rochester, Dept. of Physics and Astronomy, Rochester NY 14627, USA}

\address[usd]{University of South Dakota, Dept. of Physics, 414E Clark St., Vermillion SD 57069, USA}

\address[yale]{Yale University, Dept. of Physics, 217 Prospect St., New Haven CT 06511, USA}

\begin{abstract}
Results are presented from radioactivity screening of two models of photomultiplier tubes designed for use in current and future liquid xenon experiments. The Hamamatsu 5.6~cm diameter R8778 PMT, used in the LUX dark matter experiment, has yielded a positive detection of four common radioactive isotopes: $^{238}$U, $^{232}$Th, $^{40}$K, and $^{60}$Co. Screening of LUX materials has rendered backgrounds from other detector materials subdominant to the R8778 contribution. A prototype Hamamatsu 7.6~cm diameter R11410~MOD PMT has also been screened, with benchmark isotope counts measured at $<$0.4~$^{238}$U~/~ $<$0.3~$^{232}$Th~/~$<$8.3~$^{40}$K~/~$2.0\pm0.2$~$^{60}$Co~mBq/PMT. This represents a large reduction, equal to a change of $\times\frac{1}{24}$~$^{238}$U~/~$\times\frac{1}{9}$~$^{232}$Th~/~$\times\frac{1}{8}$~$^{40}$K per PMT, between R8778 and R11410~MOD, concurrent with a doubling of the photocathode surface area (4.5~cm to 6.4~cm diameter). $^{60}$Co measurements are comparable between the PMTs, but can be significantly reduced in future R11410~MOD units through further material selection. Assuming PMT activity equal to the measured 90\% upper limits, Monte Carlo estimates indicate that replacement of R8778 PMTs with R11410~MOD PMTs will change LUX PMT electron recoil background contributions by a factor of $\times\frac{1}{25}$ after further material selection for $^{60}$Co reduction, and nuclear recoil backgrounds by a factor of $\times\frac{1}{36}$. The strong reduction in backgrounds below the measured R8778 levels makes the R11410~MOD a very competitive technology for use in large-scale liquid xenon detectors.
\end{abstract}

\maketitle


\section{Introduction}

Photomultiplier tubes (PMTs) are commonly used for detection of photon signals in scintillation detectors. In the case of low-background detectors, the proximity of the PMTs to the active scintillation region can lead to a background dominated by PMT radioactivity. Understanding PMT radioactivity and selecting PMTs based on radiopure material composition is therefore of primary importance for experiments making use of low-background detectors, as the use of ultra-low background PMTs can greatly extend background-free running times.

The LUX dark matter direct detection experiment uses 350~kg of liquid xenon for the detection of recoils resulting from the scattering of Weakly Interacting Massive Particles (WIMPs) \cite{Akerib2011}. The experiment uses 122 Hamamatsu R8778 PMTs directly above and below the active region to detect scintillation light from xenon nuclear recoils \cite{R8778Datasheet}. R8778 PMTs are well suited for use in liquid xenon detectors. Xenon scintillation light is narrowly peaked at 178~nm \cite{jortner:4250,Doke2002JJAP}; the LUX R8778 PMTs feature a measured average 33\% quantum efficiency and 90\% collection efficiency at this wavelength. They offer single-photon sensitivity with a resolution of 30\%. Their 4.5~cm diameter photocathode allows for high surface area coverage per tube compared to PMTs used in previous xenon dark matter detectors, such as the 2.2~cm R8520 \cite{Aprile2011,Aprile:2011dd,R8520-data-sheet}. The R8778 was also specifically designed to operate at liquid xenon temperatures, in the range 165-180~K. Extensive testing has been performed on the LUX PMTs, and their behavior in liquid xenon is well understood.

Through extensive screening and material selection, it has been possible to render the background contributions from other LUX construction materials subdominant to that of the PMTs \cite{Akerib2011}. The PMTs contribute backgrounds in the form of gamma rays and neutrons produced by radioactive decay, as well as neutrons from ($\alpha$,n) reactions in PMT materials. Neutron backgrounds are of particular importance to dark matter detectors due to their similarity to a WIMP nuclear recoil signal. Precise measurements of all radioactive isotopes must be made to ascertain the impact of the PMTs on LUX backgrounds.

LUX has initiated a development program in conjunction with Hamamatsu to identify candidate PMTs for use in future dark matter detectors. The LUX-ZEPLIN (LZ) liquid xenon detector will be a factor of $\times$20 larger in mass than LUX, with corresponding surface area increase of $\times$24 \cite{LZProposal2012}. The PMTs used in these detectors will ideally be larger than the R8778, and offer lower radioactive backgrounds per unit area.

\begin{figure}
\begin{centering}
\includegraphics[width=1\columnwidth]{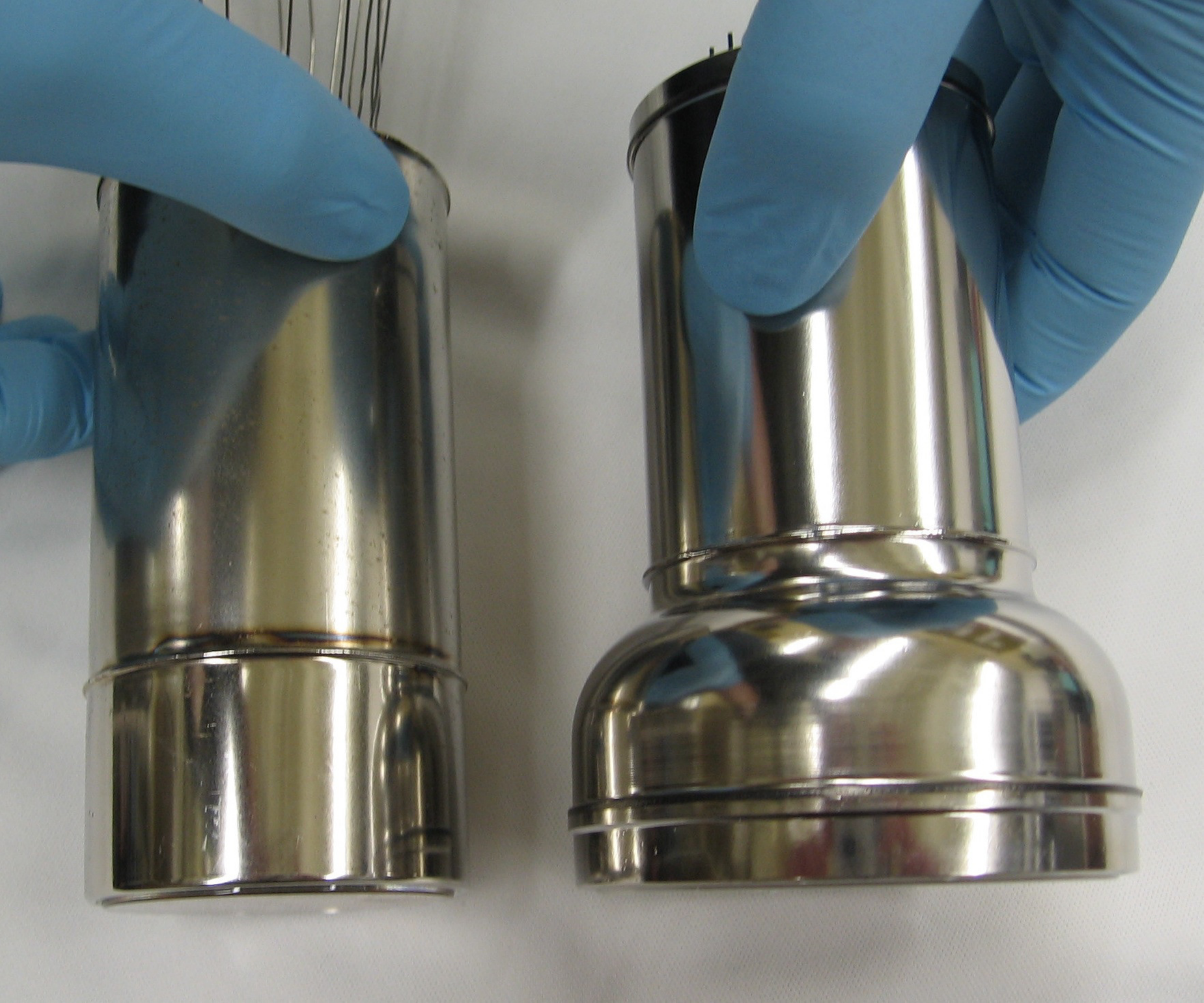}
\par\end{centering}
\caption{The Hamamatsu R8778 PMT, used in the LUX experiment (left), and the R11410~MOD PMT, a candidate for use in the LZ detector (right). The R11410~MOD features a 6.4~cm diameter photocathode, offering twice the surface area of the 4.5~cm diameter R8778 photocathode.}
\label{fig:R8778-R11410MOD-size-comparison}
\end{figure}

The development program has resulted in the creation of several large-area photocathode PMTs, notably the 7.6~cm diameter R11065 and its low-background variant, the R11410~MOD \cite{R11065,R11410MODDatasheet}. The R11065 was tested at Brown University and was found to yield an identical response to xenon scintillation light as the R8778, with gains measured up to $10^7$ at 1500~V and a single photoelectron resolution of 32\% while immersed in liquid xenon at 175~K. The R11410~MOD PMT has been selected as a candidate for use in the LZ program. The R11410~MOD and R8778 PMTs are shown for comparison in Fig.~\ref{fig:R8778-R11410MOD-size-comparison}.

\section{\label{sec:Counting-Results}Counting Results}

\subsection{\label{sub:SOLO}SOLO}

\begin{figure}
\includegraphics[width=1\columnwidth]{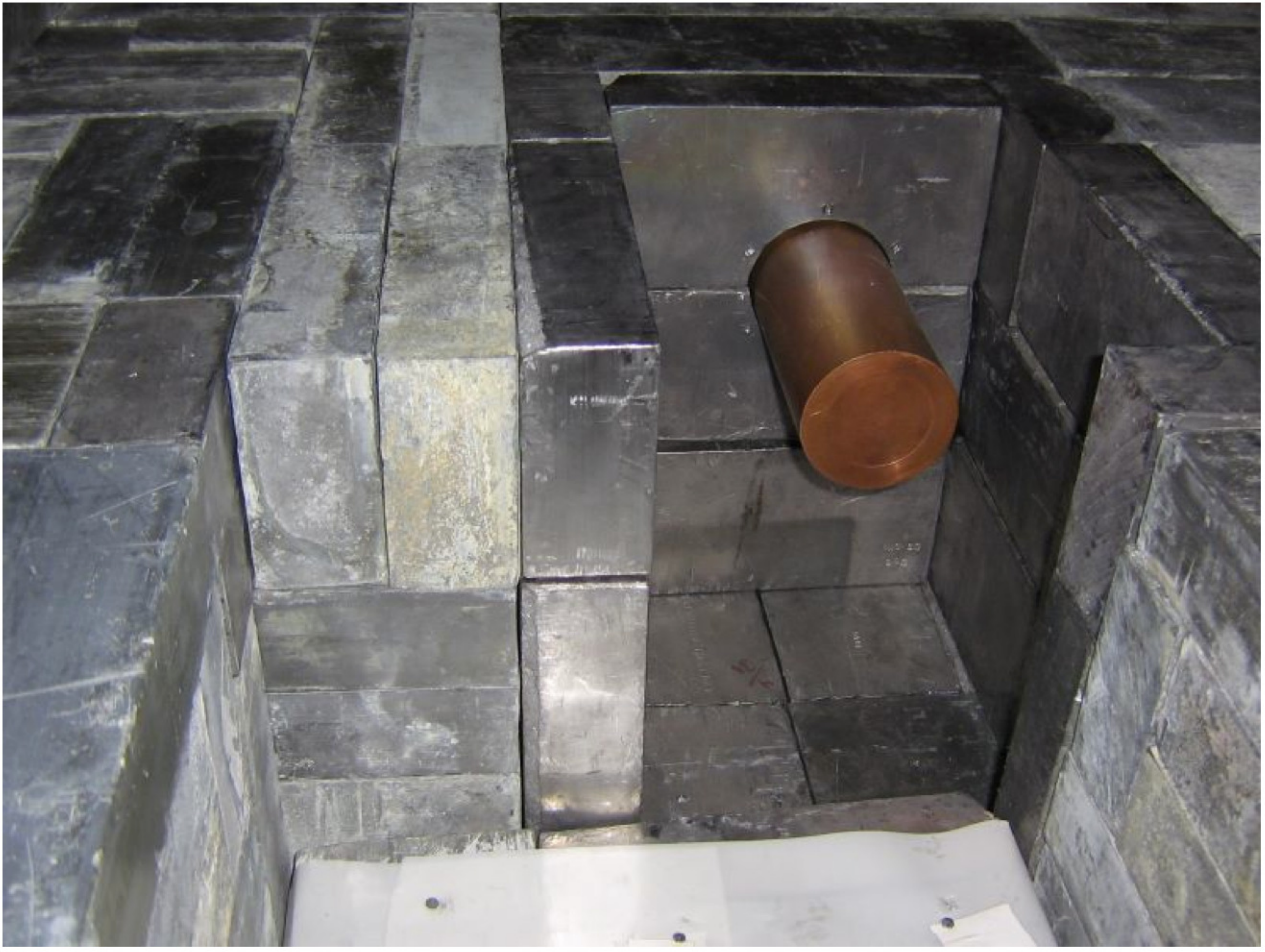}
\caption{The open SOLO chamber. The 0.6~kg high-purity germanium detector is housed in a 1.5~mm thick copper shield. The chamber is lined with >30~cm of lead shielding. The inner 5~cm lead layer was selected for its measured $^{210}$Pb content ($<$50~mBq/kg).}
\label{fig:SOLO-chamber-open}
\end{figure}

The Soudan Low-Background Counting Facility (SOLO) is located at the Soudan Underground Laboratory, at a depth of 2.0~km.w.e \cite{attisha:75}. The SOLO geometry is shown in Fig.~\ref{fig:SOLO-chamber-open}. The counting chamber uses the ``Diode-M'' detector, a 0.6~kg high-purity germanium detector housed in a 0.15~cm thick copper shield. The 20~cm $\times$ 25~cm $\times$ 30~cm chamber is enclosed with a minimum of 30~cm lead shielding on every side. The inner 5~cm lining of the chamber is comprised of ancient lead, with $^{210}$Pb activity measured below 50~mBq/kg. A mylar shell and 2.5~slpm nitrogen gas purge are used to eliminate gaseous radon from the chamber. Detector output is shaped with a 2~$\mu$s shaping time, using a Spectroscopy Amplifier, before digitization at a dedicated acquisition computer. Energy bin counts are written to data files every four hours for analysis.

SOLO has been used since 2003 for screening and cataloging of low-background materials for use in liquid xenon detectors. SOLO was used in the screening program for XENON10 construction materials, and has been heavily engaged in the LUX material screening program since 2007 \cite{Akerib2011,Aprile2011,Angle:2007uj,deViveiros2009thesis}.

\begin{figure}
\includegraphics[width=1\columnwidth]{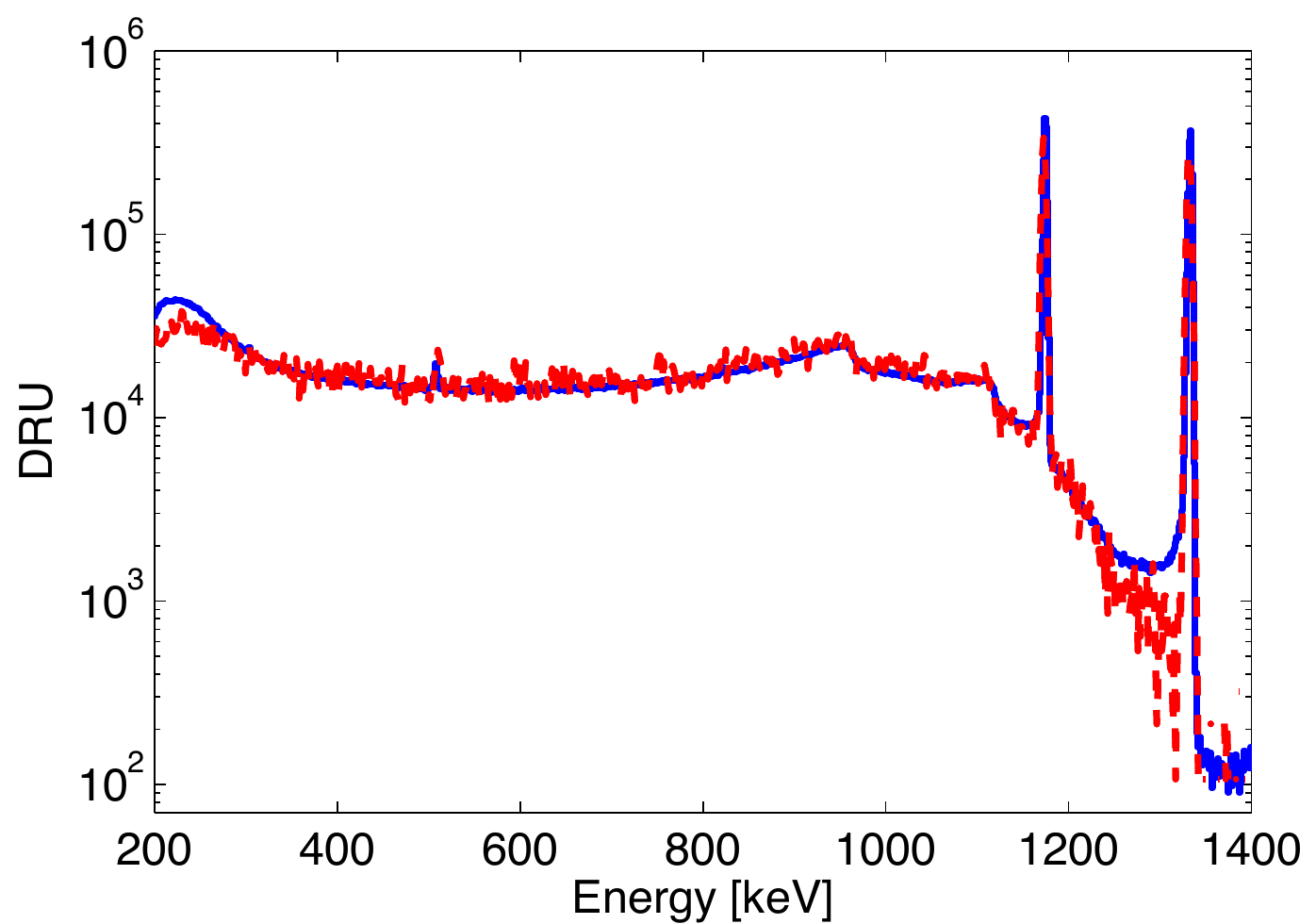}
\caption{Comparison of calibration Monte Carlo output (solid) with data from a $^{60}$Co source (dashed). Calibration Monte Carlo includes the detailed SOLO chamber and detector geometries, as well as faithful reconstruction of the source geometry to account for finite sample size and self-shielding.}
\label{fig:SOLO-MC-fit}
\end{figure}

Detector sensitivity calibrations are accomplished using Geant4 Monte Carlo simulations. The geometry of the entire SOLO detector and chamber are reproduced, as well as the geometry and location of the counted sample. Emission spectra for each screened isotope, including the full decay chains of $^{238}$U and $^{232}$Th, are reproduced. These simulations account for effects including energy-dependent detection efficiency, sample location, sample size, and sample self-shielding. The simulations produce scaling factors between detector counts and source emission intensity for each gamma line, referred to as the detection efficiency $\varepsilon$. Analytic estimates of $\varepsilon$ can be generated to cross-check simulation results. However, it is very difficult to include the effects of finite sample size and sample self-shielding in these estimates, whereas those effects are automatically included in the Monte Carlo analysis. A fit to a 1~$\mu$Ci $^{60}$Co source is shown in Fig.~\ref{fig:SOLO-MC-fit}.

\begin{figure}
\includegraphics[width=1\columnwidth]{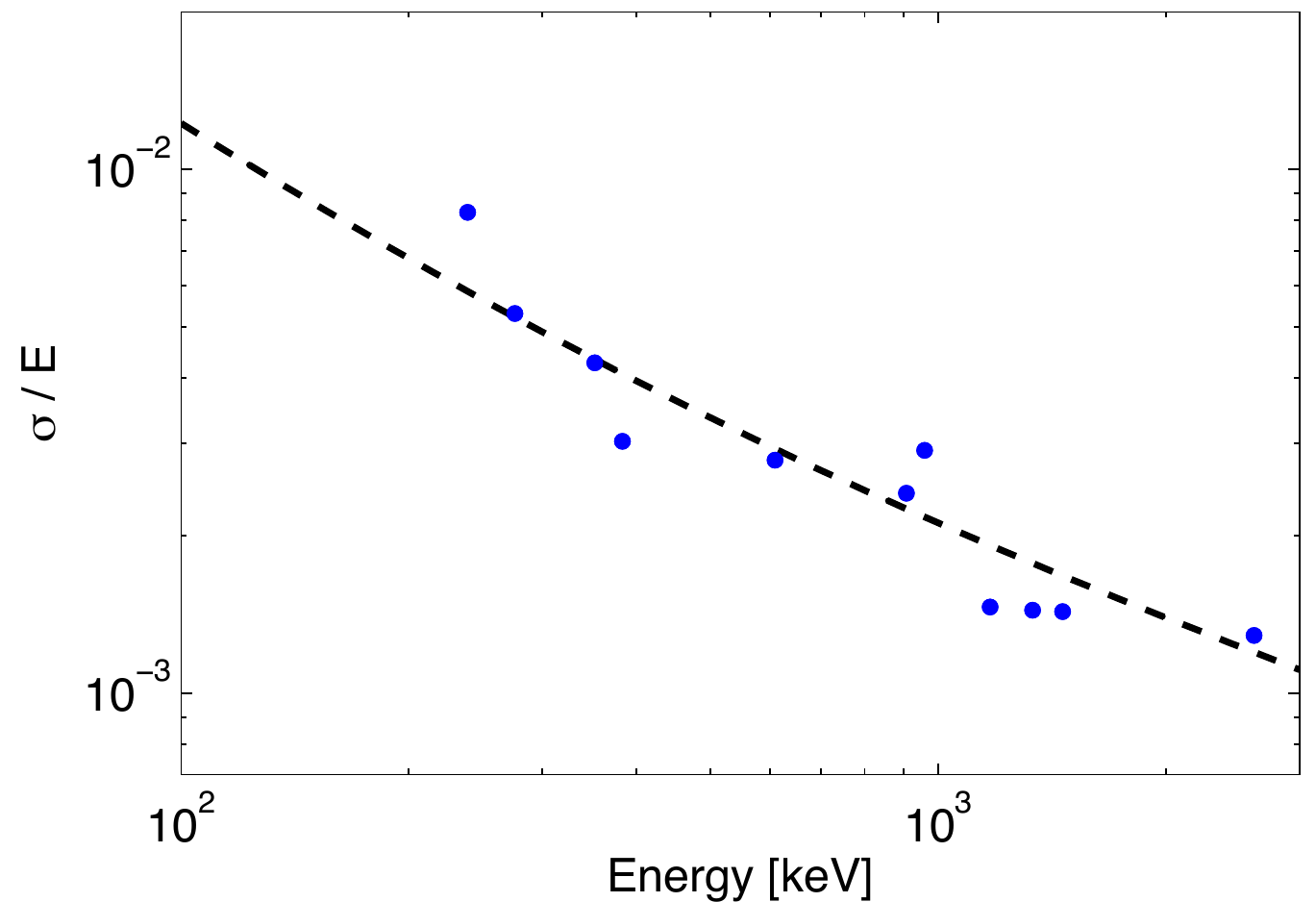}
\caption{SOLO measured $\sigma/\mu$ energy resolution as a function of energy. The measured points are characterized by the curve $\sigma^2=0.0033E+1.2$.}
\label{fig:SOLO-res}
\end{figure}

\begin{figure}
\includegraphics[width=1\columnwidth]{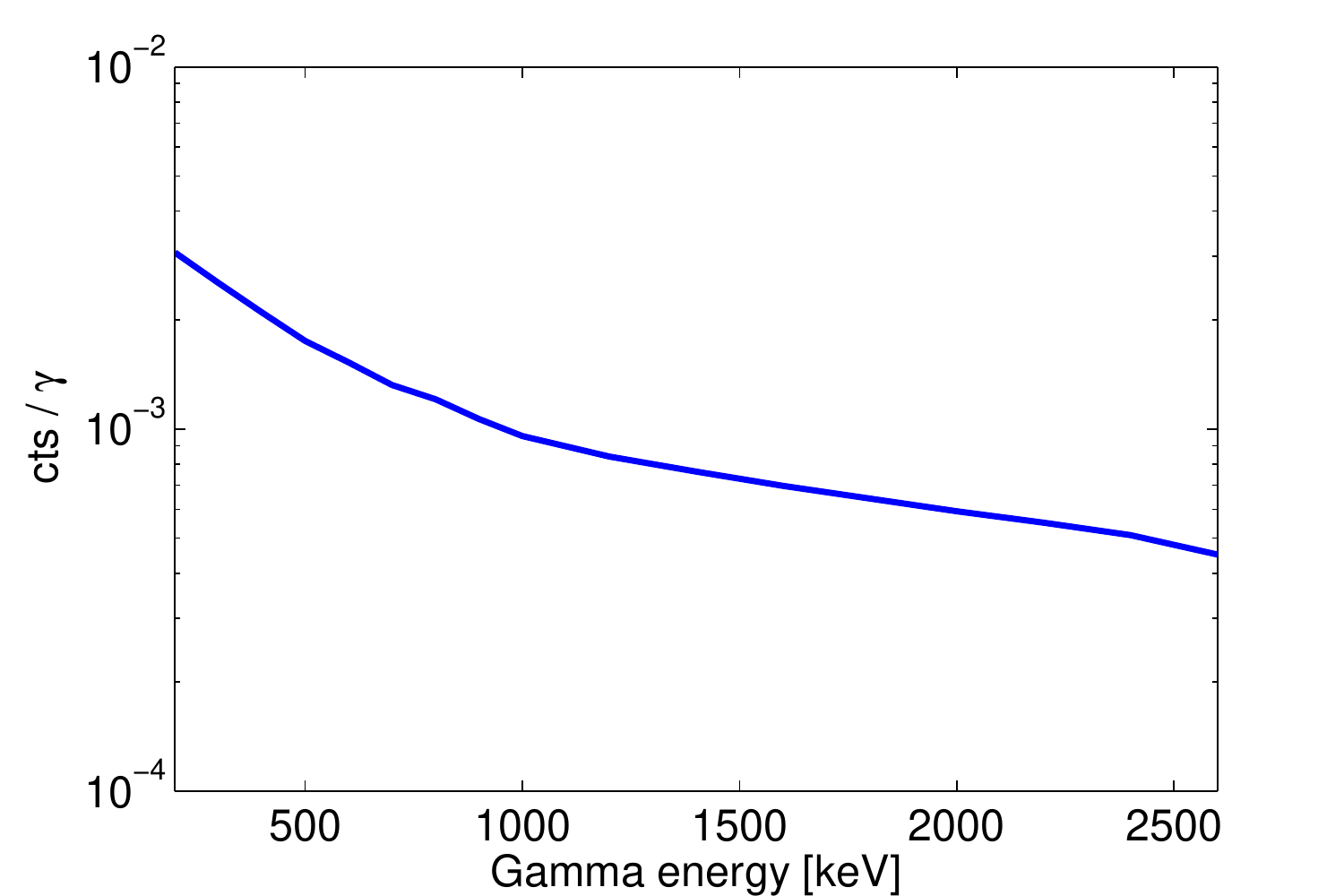}
\caption{Detection efficiency as a function of energy for a calibration point source, obtained from calibration Monte Carlo.}
\label{fig:SOLO-eff}
\end{figure}

A typical background spectrum with the chamber empty is shown in Fig.~\ref{fig:SOLO-PMT-spectra} (black) for 21 live~days. Features are seen at 662~keV and 1461~keV, corresponding to gamma lines from $^{137}$Cs and $^{40}$K, respectively, and are used to verify energy bin calibration in conjunction with the 46.5~keV line from $^{210}$Pb decay. Background baseline rise below 1~MeV is caused primarily by Bremsstrahlung radiation from $^{210}$Pb beta decay in the inner lead shielding layers. Initial detector calibrations were performed using several samples with known contamination. Backgrounds are remeasured periodically to check for chamber contamination or radon leakage, and are used to periodically verify calibrations. The measured peak energy resolution as a function of energy is shown in Fig.~\ref{fig:SOLO-res}. Peak sensitivity as a function of energy for a point calibration source at the bottom of the SOLO chamber, estimated from Monte Carlo, is shown in Fig.~\ref{fig:SOLO-eff}.

\subsection{PMT Activity Measurements}

\subsubsection{R8778}

A total of 20~R8778 PMTs were screened at SOLO in order to yield high-statistics measurements of their radioactivity. Ten R8778 PMTs, purchased for initial testing at Brown University, were screened in two separate batches of five. After the purchase of the 122 PMTs for LUX, two additional batches of five LUX PMTs were counted. Average results are listed in Table~\ref{tab:PMT-screening-results}, with the screening spectrum from one of the batches of five R8778 PMTs shown in Fig.~\ref{fig:SOLO-PMT-spectra}. Counting results for $^{238}$U / $^{232}$Th / $^{60}$Co varied between batches by less than 20\%. $^{40}$K was reduced between the advance PMT set and the LUX PMTs by 30\% after iteration in material selection by Hamamatsu. Results indicate the presence of benchmark isotopes $^{238}$U, $^{232}$Th, $^{40}$K and $^{60}$Co in amounts of $9.5\pm0.6$, $2.7\pm0.3$, $66\pm6$, and $2.6\pm0.2$ mBq/PMT, respectively. The measured values represent an overall gamma ray emission rate per PMT that is a factor of 3 below the conservative estimate originally used in the LUX proposal, which used the manufacturer-supplied upper limits of $<$18~$^{238}$U / $<$17~$^{232}$Th / $<$30~$^{40}$K / $<$8~$^{60}$Co mBq/PMT.

\subsubsection{R11410~MOD}

An R11410~MOD PMT mechanical sample was screened at SOLO for a period of 19 live~days. This PMT was produced after a thorough material screening program by the manufacturer, with strong reductions in benchmark isotopes expected. Placement of the R11410~MOD was optimized within the chamber, and the counting time was extended a factor $\times$2 beyond the typical time for the R8778 batches, in order to maximize sensitivity for the single sample.

\begin{table*}
\centering
\small
\begin{tabular}{| c | c | c | c | c | c | c | c |}
\hline 
\textbf{PMT}	& \multicolumn{5}{c |}{\textbf{Activity [mBq/PMT]}}	\\
			& \textbf{$^{238}$U ($^{234\text{m}}$Pa)} & \textbf{$^{238}$U ($^{226}$Ra)} & \textbf{$^{232}$Th ($^{228}$Ra)} & \textbf{$^{40}$K} & \textbf{$^{60}$Co}	\\
\hline 
\hline 
R8778 & $<$22 & $9.5\pm0.6$ & $2.7\pm0.3$ & $66\pm6$ & $2.6\pm0.2$	\\
R11410~MOD & $<$6.0 & $<$0.4 & $<$0.3 & $<$8.3 & $2.0\pm0.2$	\\
\hline
\end{tabular}
\vskip 8pt
\caption{R8778 and R11410~MOD radioactivity screening results, obtained at SOLO. All PMTs were screened for four benchmark isotopes identified by the factory: $^{238}$U, $^{232}$Th, $^{40}$K, and $^{60}$Co. R8778~PMTs were counted in four batches of 5~units, with an average 9~live~days per batch. Averaged results are listed. The R11410~MOD sample was counted for 19~live~days. Separate columns are provided for completeness for ``early'' and ``late'' $^{238}$U chain measurements, allowing for potential equilibrium breakage primarily due to $^{226}$Ra solubility in water; however, it should be noted that there is no \emph{a priori} reason to assume an equilibrium breakage. Errors are statistical and quoted at $\pm$1$\sigma$. Upper limits are given at 90\% confidence level. Results are normalized per PMT for comparison.}
\label{tab:PMT-screening-results}
\end{table*}

\begin{figure*}
\begin{centering}
\includegraphics[width=1\columnwidth]{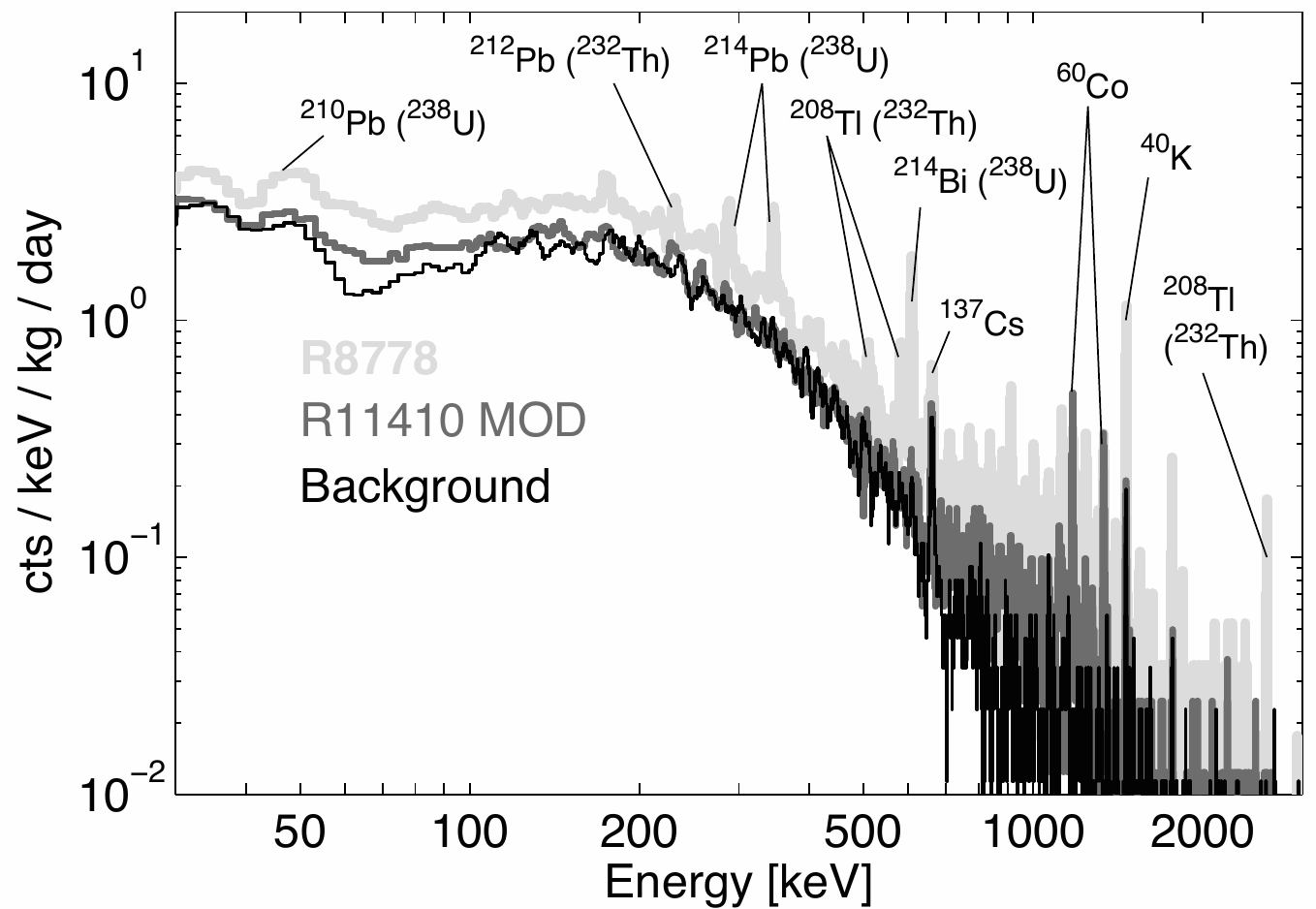}
\par\end{centering}
\caption{SOLO counting spectra for R8778 (14~live~days, blue) and R11410~MOD (19~live~days, red) PMTs, superimposed with a sample background run (21~live~days, black). $^{238}$U is primarily identified by lines from late-chain daughters at 295, 352, and 609~keV. $^{232}$Th is identified by lines at 239, 511, 582, and 2614~keV. $^{40}$K is identified by a single line at 1460~keV. $^{60}$Co yields twin peaks at 1173 and 1333~keV. Strong lines in the R8778 spectrum indicate the presence of $^{238}$U, $^{232}$Th, $^{40}$K and $^{60}$Co radionuclides. The reduced activity of the R11410~MOD is readily apparent from the lack of distinct features in comparison with the R8778, with the only activity significantly above background from the presence of $^{60}$Co. Lines at 46.5~keV ($^{210}$Pb) and 662~keV ($^{137}$Cs) are used to periodically verify detector calibration.}
\label{fig:SOLO-PMT-spectra}
\end{figure*}

The R11410~MOD showed a very significant reduction in $^{238}$U, $^{232}$Th and $^{40}$K content compared to R8778 counting results, and the signal for these isotopes is consistent with background. The only positive signal was found at 1173 and 1333~keV, consistent with the presence of $^{60}$Co in amounts comparable to those measured in the R8778 PMTs. Further material selection can significantly reduce the presence of $^{60}$Co in R11410~MOD units, with a goal of <0.2~mBq/PMT for complete subdominance. R11410~MOD counting results are summarized alongside R8778 results in Table~\ref{tab:PMT-screening-results}, and the screening spectrum is overlaid with the R8778 spectrum and background in Fig.~\ref{fig:SOLO-PMT-spectra}.

Material selection has been carried out by Hamamatsu, with support from SOLO used for screening several different material samples. Additional refinements continue to be made on the R11410 models; SOLO will be used to screen large material batches of all major components, in an effort to identify any remaining high-activity materials. SOLO will also be used to screen several batches of R11410-20 PMTs, the newest iteration of the low-background R11410 line, in support of the LZ experiment.

The number of PMTs used in an experiment is inversely proportional to the photocathode area per PMT. Thus a proper comparison of PMT activity must also normalize radioactivity by photocathode area. Screening results normalized per unit photocathode area are listed for R8778 and R11410~MOD PMTs, as well as for 17 batches of R8520 PMTs counted at the University of Zurich and Laboratori Nazionali del Gran Sasso (LNGS) for the XENON collaboration \cite{Kish2011}, in Table~\ref{tab:PMT-screening-normalized}. The R11410~MOD offers by far the lowest activity per unit area, with upper limit activities a factor of $\times\frac{1}{3}$ / $\times\frac{1}{4}$ / $\times\frac{1}{10}$ that of the R8520 positive measurements in $^{238}$U / $^{232}$Th / $^{40}$K, respectively. Activity per unit area for $^{60}$Co is measured to be a factor of $\times\frac{1}{3}$ below that of the R8520, and would drop dramatically after material selection for future R11410~MOD units.

\begin{table*}
\centering
\small
\begin{tabular}{| c | c | c | c | c | c |}
\hline 
\textbf{PMT}	& \multicolumn{5}{c |}{\textbf{Photocathode-Normalized Activity [mBq/cm$^2$]}}	\\
			& \textbf{$^{238}$U ($^{234\text{m}}$Pa)} & \textbf{$^{238}$U ($^{226}$Ra)} & \textbf{$^{232}$Th ($^{228}$Ra)} & \textbf{$^{40}$K} & \textbf{$^{60}$Co}		\\
\hline 
\hline 
R11410~MOD (this work)			& $<$0.19		& $<$0.013 		& $<$0.009 		& $<$0.26 		& 0.063$\pm$0.0063	\\
R8778 (this work)				& $<$1.4		& 0.59$\pm$0.038 	& 0.17$\pm$0.019 	& 4.1$\pm$0.13 	& 0.16$\pm$0.0063		\\
\hline
R8520 (XENON10)  \cite{deViveiros2009thesis}	& 			& 0.03			& 0.03			& 2				& 0.4					\\
R8520 (XENON100) \cite{Kish2011} 			& $<$3.4		& 0.038$\pm$0.002 	& 0.035$\pm$0.002 	& 2.3$\pm$0.07 	& 0.18$\pm$0.005		\\
R11410~MOD \cite{Kish2011}		& $<$3.0		& $<$0.075		& $<$0.12			& 0.40$\pm$0.1	& 0.11$\pm$0.02		\\
\hline 
\end{tabular}
\vskip 8pt
\caption{Counting results from Table~1, renormalized by PMT photocathode area. Based on work in this paper, the R11410~MOD offers the lowest radioactivity per PMT per unit photocathode area for all benchmark isotopes. In particular, ratios of R11410~MOD upper limits to R8778 activities are $\times1/45$~$^{238}$U / $\times1/19$~$^{232}$Th / $\times1/16$~$^{40}$K. The ratio for $^{60}$Co, detected in both PMTs, is $\times1/3$, and can be significantly reduced with further material screening as discussed in the text. The very low normalized ratios make the R11410~MOD the top candidate for use in the large-area tonne-scale LZ detectors. Included for comparison are counting results for the R8520 PMTs used in the XENON10 \cite{deViveiros2009thesis} and XENON100 \cite{Kish2011} experiments, as well as counting results for a second R11410~MOD PMT presented in \cite{Kish2011}.}
\label{tab:PMT-screening-normalized}
\end{table*}

\subsection{Counting Methodology}

Positive signatures of radioactive isotopes are identified through two high-branching ratio energy lines, when available. For measurement of the $^{238}$U chain, lines at 353~keV and 609~keV are used, corresponding to the decay of $^{214}$Pb and $^{214}$Bi respectively. The $^{232}$Th chain is identified by lines at 511~keV and 583~keV from $^{208}$Tl decay. $^{60}$Co is detected through peaks at 1173~keV and 1333~keV. $^{40}$K offers only a single gamma line, at 1461~keV.

It has been noted that $^{238}$U decay chain secular equilibrium can be broken during certain manufacturing processes due to the solubility of $^{226}$Ra in water. For this reason, a more direct estimate of $^{238}$U concentration is made from measurement of the 1001~keV line from $^{234\text{m}}$Pa, several steps above $^{226}$Ra in the decay chain. The branching ratio for the 1001~keV line is a factor of $\times\frac{1}{62}$ that of the branching ratio for the 353~keV line, and thus produces correspondingly weaker upper limits. It should be noted that there is no \emph{a priori} reason to assume a break in equilibrium, unless a positive identification of the 1001~keV line is made. It should also be noted that the most relevant, high branching ratio gammas are generated in the lower part of the $^{238}$U chain, and this measurement is of the most relevance for the estimation of electromagnetic backgrounds.

Upper limit estimates are always given at 90\% confidence level, the standard used during LUX material counting. Quoted errors are statistical. Systematic errors are contributed primarily by uncertainty in sample placement within the chamber and in sample self-shielding, and are conservatively estimated to be at the level of $\pm$20\%.

Upper limits are calculated assuming a Poisson-distributed signal. The measured value $N$, in counts, is defined as $N\equiv S+B$, where $S$ is a positive signal and $B$ is the separately-measured background, and either $N$ or $B$ is scaled down such that the number of live~days for the two quantities is equal. As $S$ depends on the efficiency of gamma detection, it is defined as $S\equiv\varepsilon\alpha$, where $\alpha$ is the source radioactivity in units of mBq and $\varepsilon$ is the detection efficiency already discussed in Sec.~\ref{sub:SOLO}. $\varepsilon$ is measured in counts/mBq, and is obtained from Monte Carlo calibrations, detector mass, and counting livetime. For a given confidence level $\mbox{CL}$ (equal to 0.9 for all SOLO results), the following integral is numerically evaluated to determine the parameter $\alpha'$:

\[
\mbox{CL}=\frac{\intop_{0}^{\alpha'}p\left(N,\varepsilon\alpha+B\right)d\alpha}{\intop_{0}^{\infty}p\left(N,\varepsilon\alpha+B\right)d\alpha}
\]

where $p(N,\lambda)$ is the Poisson distribution function for measured value $N$ and expectation $\lambda$. The value of $\alpha'$ then represents the upper limit source activity at confidence level $\mbox{CL}$.

\section{Impact on Dark Matter Backgrounds}

\subsection{LUX Backgrounds}

Counting results indicating a change of $\times\frac{1}{24}$ in $^{238}$U and $\times\frac{1}{9}$ in $^{232}$Th between R8778~PMTs and R11410~MOD~PMTs are of particular importance, as these isotopes are primary contributors of neutron backgrounds in dark matter detectors. $^{238}$U and $^{232}$Th both create neutrons through ($\alpha$,n) reactions in the surrounding PMT materials. $^{238}$U can also undergo spontaneous fission, directly producing neutrons with a contribution similar to that of ($\alpha$,n). Neutron rates from ($\alpha$,n) are calculated using the Neutron Yield Tool developed by LUX collaborators at the University of South Dakota \cite{neutronyield}, based upon work detailed in \cite{Mei2008ir}. The $^{238}$U concentration within the PMT is assumed to be equally distributed amongst all PMT components for the calculations. Neutron yields are calculated separately for each chemical compound in each component in order to more accurately account for material stopping power effects.

The ($\alpha$,n) rates for the R8778 PMTs, summed for all components, are calculated as 0.60~n/PMT/yr and 0.21~n/PMT/yr for $^{238}$U and $^{232}$Th, respectively. The spontaneous fission rate of $^{238}$U is calculated as $1.13\times10^{-6}$~n/decay \cite{Sultis2002fundamentals}, yielding an additional 0.33~n/PMT/yr. This totals to 1.2~n/PMT/yr. The R11410~MOD counting upper limits represent a neutron emission rate change of $\times\frac{1}{18}$, or an upper limit of 0.06~n/PMT/yr.

The LUX experiment could potentially benefit from the use of 62 R11410~MOD PMTs, replacing the 122 R8778 PMTs currently in use. Neutron backgrounds are reduced in direct proportion to the product of the number of neutrons produced per PMT per unit time and the number of PMTs used in the experiment. In the case of LUX, the neutron background reduction factor after the R8778~PMTs are replaced by R11410~MOD~PMTs is then $\times\frac{1}{36}$. This represents a total of 3.0 neutrons produced from the PMT arrays per 300~live~day running time; after single-scatter and energy cuts, which reduce event rates by over four orders of magnitude, the neutron background expectation in the WIMP search energy region is virtually zero. 

Estimates of electron recoil (ER) background contributions from LUX PMTs have been generated using the LUXSim Monte Carlo simulation package developed by LUX collaborators \cite{LUXSim}. LUXSim results indicate that the use of the R11410~MOD PMT arrays, conservatively assuming activity levels equal to their measured 90\% upper limit values, would change LUX ER backgrounds by a factor of $\times\frac{1}{10}$. This assumes a WIMP search nuclear recoil (NR) energy range of 5-25~keV$_{\mbox{nr}}$, matching that used in the LUX proposal. It is important to note that this estimate uses the current measurement of R11410~MOD $^{60}$Co, which is expected to be significantly reduced in future production models. With a reduction of $^{60}$Co to levels of $<$0.2~mBq/PMT, the background reduction factor would become $\times\frac{1}{25}$, again making the conservative assumption of $^{238}$U / $^{232}$Th / $^{40}$K levels equal to their measured 90\% upper limits. The strong reduction in both ER and NR backgrounds indicates that the R11410~MOD PMTs can potentially reduce the overall LUX PMT background contribution below that of other construction materials used in the experiment, creating an overall change in predicted LUX backgrounds of nearly $\times\frac{1}{10}$.

\begin{figure*}
\begin{centering}
\includegraphics[width=0.49\columnwidth]{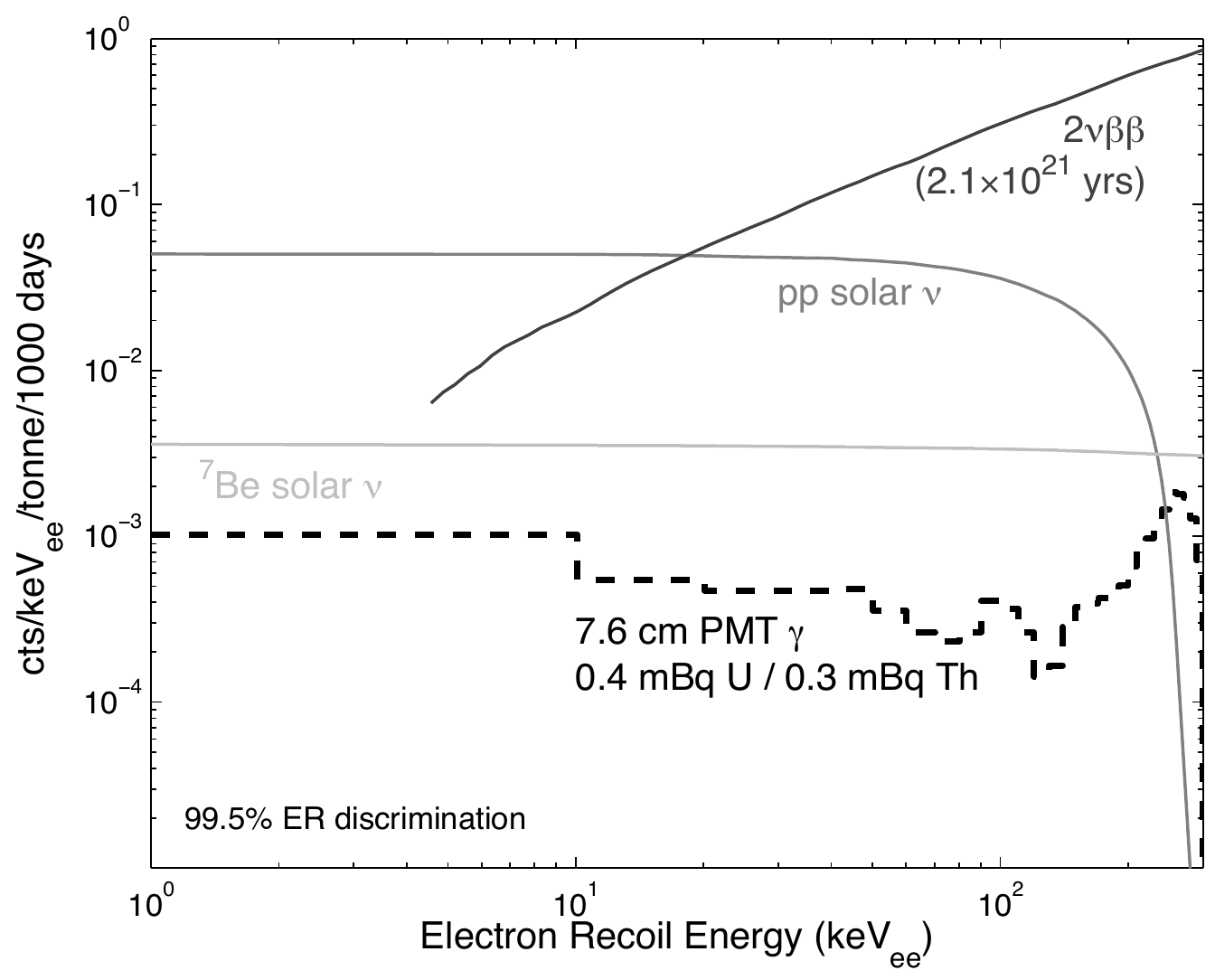}
\includegraphics[width=0.49\columnwidth]{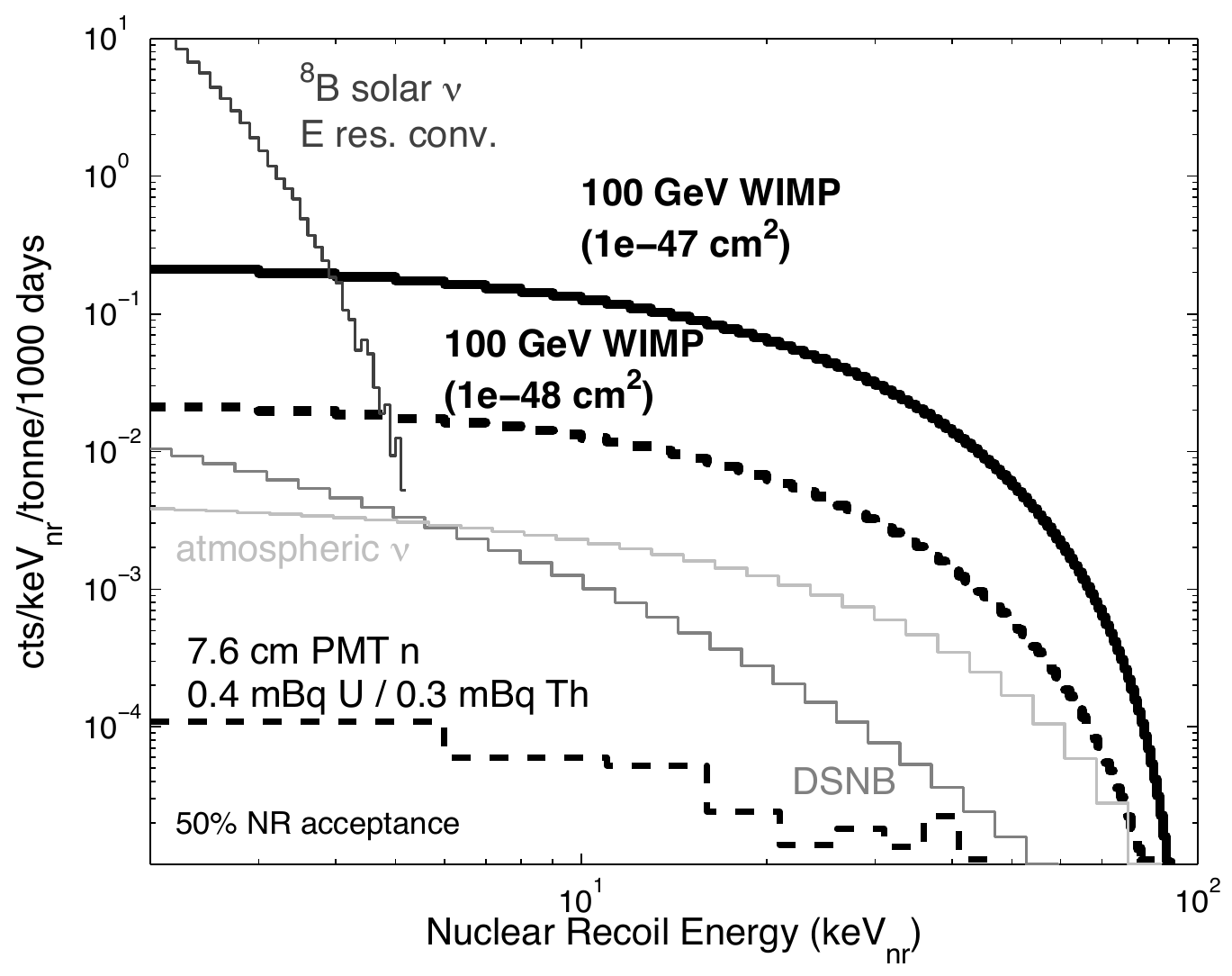}
\par\end{centering}

\caption{(Left) LZ ER backgrounds from neutrino scattering and $^{136}$Xe two-neutrino double-beta decay, assuming a $2.1\times10^{21}$~yr lifetime as recently reported in \cite{Ackerman2011}. PMT $^{238}$U and $^{232}$Th ER background projections are overlaid for the case of R11410~MOD PMTs. A conservative 99.5\% rejection factor is applied for neutrino and gamma spectra. (Right) LZ neutrino coherent scattering backgrounds adapted from \cite{Strigari:2009bq}. Overlaid are neutron background spectra for both ($\alpha$,n) and fission contributions from the PMTs. 50\% NR acceptance is assumed.}
\label{fig:LZ-nu-backgrounds}
\end{figure*}

\subsection{LZ Backgrounds}

It is interesting to note that no further reduction in PMT radioactivity beyond the R11410~MOD is useful for the LZ experiment. In particular, LZ faces a background ``floor'' from neutrino scattering, causing ER events as well as NR events from coherent neutrino scattering \cite{Strigari:2009bq}. The spectra for both ER and NR backgrounds are shown in Fig.~\ref{fig:LZ-nu-backgrounds}.

ER backgrounds are predicted to be dominated by scattering from solar neutrinos from the p-p chain, even with the use of PMTs equivalent to the LUX R8778. With the use of R11410~MOD PMTs in the tonne-scale experiment, the PMT NR background contribution can be reduced below the level of coherent neutrino scattering from atmospheric and diffuse supernova background neutrinos above 5~keV. This will facilitate a direct measurement of these scattering spectra.

It can be noted that a PMT with an activity per surface area corresponding to R11410~MOD counting results presented in \cite{Kish2011} (Table \ref{tab:PMT-screening-normalized}) would increase ER backgrounds by a factor of $\times$7 and NR backgrounds by a factor $\times$10. The PMT background contributions will then be a factor $\times$1/7 that of neutrino ER contributions, and in a range $\times$1/10 - $\times$1/4 that of neutrino NR contributions. Experiment backgrounds are negligibly affected by this increase, as are neutrino spectra measurements.

\section{Conclusion}

The Hamamatsu R8778 and R11410~MOD PMTs are both well suited for use in liquid xenon scintillation detectors. The radioactive isotope content of both PMT models has been assessed. Counting results for the R11410~MOD PMT yield a combined $^{238}$U + $^{232}$Th measurement below 0.7~mBq/PMT.

Background simulations have been used to estimate the potential reduction in LUX backgrounds from replacement of the current LUX PMTs with R11410~MOD PMTs. Simulation results indicate that changes of $\times\frac{1}{25}$ in LUX PMT electron recoil backgrounds and $\times\frac{1}{36}$ in neutron recoil backgrounds are expected, conservatively assuming PMT activities equal to the 90\% upper limits found for all benchmark isotopes, and after reduction of $^{60}$Co limits to <0.2~mBq/PMT through material selection in future R11410~MOD units. At this point, LUX experiment backgrounds would be dominated by other construction materials, and would be well below the level of 0.1~WIMP-like events during the 300~live~day running time of the experiment. The use of these PMTs in the tonne-scale LZ detector will reduce the PMT background contribution below the irreducible contributions from solar and atmospheric neutrinos; no further reductions in radioactivity beyond the measurements for the R11410~MOD are necessary for these next-generation experiments.

The counting results presented here make R11410~MOD PMTs a very competitive technology for use in future large-scale xenon detectors. The counting results also offer a strong baseline for future photodetector development. The PMTs have a high average measured quantum efficiency of 33\% at xenon scintillation wavelengths, as well as signal gains up to $10^7$ at 1500~V. The use of ultra-low background PMTs such as the R11410~MOD greatly aids in the development of next-generation detectors with essentially background-free discovery potential.

\section*{Acknowledgements}

The authors are very grateful to Hamamatsu Photonics K.K. for their work in manufacturing and material selection for the PMTs tested, as well as their assistance during PMT testing and radioactivity measurements. The authors would also like to recognize the assistance of Jim Beaty and Dave Saranen at Soudan Underground Lab for their work in managing the SOLO detector and data acquisition system. This work was partially supported by the U.S. National Science Foundation under award numbers PHY-0919261 and PHY-0707051, and the U.S. Department of Energy (DOE) under award number DE-FG02-91ER40688.

\end{document}